\documentclass[aps,twocolumn,showpacs,nofootinbib]{revtex4-1}

\usepackage{color} 

\definecolor{green}{rgb}{0,.5,0}

\definecolor{red}{rgb}{1,0,0}

\usepackage{graphicx}
\usepackage[subfigure]{graphfig}
\usepackage{epsfig}
\usepackage{dcolumn}
\usepackage{amsmath}
\usepackage{multirow}
\usepackage{ulem}
\usepackage{float}
\usepackage{diagbox}

\def\bea{\begin{eqnarray}}

\def\eea{\end{eqnarray}}

\def\bal#1\eal{\begin{align}#1\end{align}}

\immediate\write18{texcount -inc -incbib 
-sum borra.tex > /tmp/wordcount.tex}

\newcommand{\lc}[1]{{\color{blue}#1}}

\begin{document}

\title{\vspace{1.0in}Distance between various discretized fermion actions}

\author{{Dian-Jun Zhao$^{1,2}$, Gen Wang$^{3,4}$, Fangcheng He$^{2}$, Luchang Jin$^{5,6}$, Peng Sun$^{7}$, Yi-Bo Yang$^{1,2,8,9}$, Kuan Zhang$^{1,2}$}
\vspace*{-0.5cm}
\begin{center}
\large{
\vspace*{0.4cm}
\includegraphics[scale=0.20]{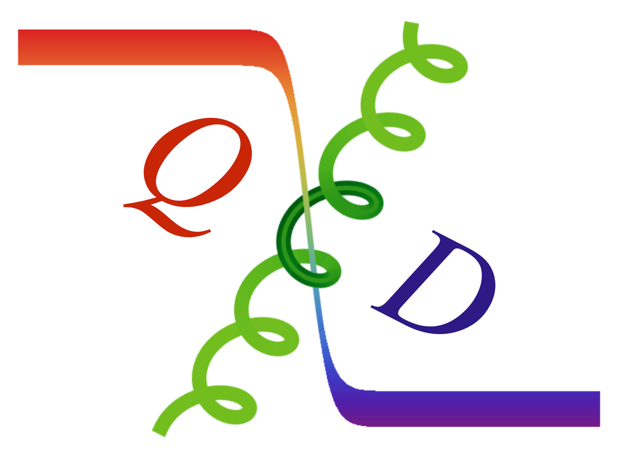}\\
\vspace*{0.4cm}
($\chi$QCD Collaboration)
}
\end{center}
}
\affiliation{
$^{1}$\mbox{University of Chinese Academy of Sciences, School of Physical Sciences, Beijing 100049, China}\\
$^{2}$\mbox{CAS Key Laboratory of Theoretical Physics, Institute of Theoretical Physics, Chinese Academy of Sciences, Beijing 100190, China}\\
$^{3}$\mbox{Department of Physics and Astronomy, University of Kentucky, Lexington, KY 40506, USA}\\
$^{4}$\mbox{Aix-Marseille Université, Université de Toulon, CNRS, CPT, Marseille, France}\\
$^{5}$\mbox{Physics Department, University of Connecticut, Storrs, CT 06269, USA}\\
$^{6}$\mbox{RIKEN-BNL Research Center, Brookhaven National Laboratory, Upton, NY 11973, USA}\\
$^{7}$\mbox{Institute of Modern Physics, Chinese Academy of Sciences, Lanzhou, 730000, China}\\
$^{8}$\mbox{School of Fundamental Physics and Mathematical Sciences, Hangzhou Institute for Advanced Study, UCAS, Hangzhou 310024, China}\\
$^{9}$\mbox{International Centre for Theoretical Physics Asia-Pacific, Beijing/Hangzhou, China}\\
}

\begin{abstract}
We present the leading order mixed-action effect $\Delta_{\rm mix}\equiv m_{\pi,{\rm vs}}^2-\frac{m_{\pi,{\rm vv}}^2+m_{\pi,{\rm ss}}^2}{2}$ using HISQ, clover or overlap valence fermion actions on gauge ensembles using various sea fermion actions across a widely-used lattice spacing range $a\in [0.04,0.19]$~fm. The results suggest that $\Delta_{\rm mix}$ decreases as the fourth order of the lattice spacing on the gauge ensembles with dynamical chiral sea fermions, such as Domain wall or HISQ fermions. When a clover sea fermion action which has explicit chiral symmetry breaking is used in the ensemble, $\Delta_{\rm mix}$ can be much larger regardless of the valence fermion action used.
\end{abstract}

\maketitle


{\it Introduction:}
Lattice provides an unique gauge invariant, non-perturbative regularization method for non-abelian gauge field theories.
However, the infamous fermion doubling problem prevents a straightforward discretization of the continuum Dirac fermion action used in 4-dimensional lattice QCD calculations.
The overlap (OV) fermion~\cite{Narayanan:1994gw,Neuberger:1997fp,Chiu:1998gp}, which satisfies the Ginsburg-Wilson relation, would be the optimal choice for the discretized Dirac operator, but requires a factor of ${\cal O}(100)$ cost of computational resources comparing to the widely used Wilson-like fermions; on the other hand, the staggered fermion and its improved versions can also provide exact chiral symmetry with a cost much lower than the Wilson-like fermion, at the expense of mixing between four equivalent ``tastes" of a given flavor. Ideally, results obtained with different fermion formulations are expected to agree in the continuum limit. But for practical lattice spacings used in the state of the art lattice calculations, it is not entirely clear to what extent they should agree~\cite{Aubin:2022hgm}.

Since, in practice, generating large ensembles using an expensive fermion action can take several months or even years, using a more expensive ``valence" fermion action on an ensemble
generated with a cheaper ``sea" fermion action has become a popular compromise in the past decade, such as in the calculations of the glue helicity~\cite{Yang:2016plb}, nucleon axial charge~\cite{Chang:2018uxx,Jang:2019vkm}, hadron vacuum polarization~\cite{Borsanyi:2020mff}, and so on. But most of the conventional lattice QCD studies still prefer a single fermion action for both the valence and sea fermions, as the ``mixed action" setup can introduce additional discretization effects as well as those (e.g., Ref.~\cite{Horsley:2004mx}) from the valence and sea fermions themselves.

From the analytical side, the mixed action partially quenched chiral perturbation theory (MAPQ$\chi$PT)~\cite{Bar:2003mh,Bar:2005tu,Chen:2007ug} suggests that the only mixed-action effect at leading order is replacing the mass squared $m_{\pi,{\rm vs}}^2$ of ``mixed-action pion" with one valence quark and one sea anti-quark into $m_{\pi,{\rm vs}}^2=\frac{m_{\pi,{\rm vv}}^2+m_{\pi,{\rm ss}}^2}{2}+\Delta^{\rm B/A}_{\rm mix}(a)$, where $m_{\pi,{\rm vv}}$ is the pion mass with two valence quarks of action A, and $m_{\pi,{\rm ss}}$ is that with two sea quarks of action B. After this replacement, standard partially quenched (PQ) $\chi$PT can be applied to study nucleon twist-2 matrix elements, the nucleon-nucleon system, neutron EDM and so on~\cite{Chen:2007ug}. Recent high-precision mixed-action lattice QCD studies with different pion masses also show that the pion form factor (which is related to the rho meson pole) can be described by PQ$\chi$PT if and only if the above mixed action replacement is applied~\cite{Wang:2020nbf}. Thus we can define an additional leading order MAPQ$\chi$PT low energy constant as
\bea\label{eq:def_mix}
\Delta^{\rm B/A}_{\rm mix}(m_{\pi,{\rm vv}},m_{\pi,{\rm ss}},a)\equiv m_{\pi,{\rm vs}}^2-\frac{m_{\pi,{\rm vv}}^2+m_{\pi,{\rm ss}}^2}{2},
\eea
and it is generally assumed that $\Delta^{\rm B/A}_{\rm mix}$ is a ${\cal O}(a^2)$ effect in most of the cases.

On the other hand, the numerical lattice QCD studies are very limited in number. Assuming $m_{\pi,{\rm vv}}=m_{\pi,{\rm ss}}\sim 300$~MeV and  $a\sim$0.09 fm, the mixed action effect will make $\delta m_{\pi}\equiv m_{\pi,{\rm vs}}-m_{\pi,{\rm ss}}$ to be 153(59) MeV for the overlap valence quark on the clover (CL, a typical Wilson-like fermion) sea~\cite{Durr:2007ef}, while $\delta m_{\pi}$ will be reduced to 30-60 MeV for the combination of the Domain wall fermion (DW, a different implementation of the overlap fermion by introducing an extra fifth dimension for the fermion) valence and staggered sea~\cite{Orginos:2007tw,Aubin:2008wk,Berkowitz:2017opd}, and it can be as small as $\sim 10$ MeV if we use the overlap valence quark on the DW sea~\cite{Lujan:2012wg}. There are also studies on other combinations, such as overlap on highly improved staggered quarks (HISQ)~\cite{Follana:2006rc} at $a\simeq 0.12$ fm~\cite{Basak:2014kma}. 

However, all of the above estimates, which are based on the imprecise calculation at $a\sim$0.1 fm, can misrepresent the
situation and render the reliability of any high-precision mixed-action calculation questionable. In this work, we present the most systematic mixed-action-effect study so far with several valence and sea fermion combinations. The results suggest that $\Delta_{\rm mix}$ decreases faster than the naive $a^2$ estimate in all the cases we studied, and at ${\cal O}(a^4)$ when the fermion action used in the sea has chiral symmetry.

{\it Methodology and setup:} The naive fermion action has the infamous fermion doubling problem, and it leads to 16 fermions as opposed to 1 in 4 dimensions. The Staggered fermion introduces a redefinition of the quark field (e.g., $\psi_{\rm st}(x)=\gamma_4^{x_4}\gamma_1^{x_1} \gamma_2^{x_2} \gamma_3^{x_3}\psi(x)$ at the site $x=\{x_1,x_2,x_3,x_4\}$ in the MILC convention) to partially fix the fermion doubling problem, and further improvement such as the use of the HISQ action~\cite{Follana:2006rc} is needed to suppress the mixing between the four residual fermions. On the other hand, the Wilson fermion avoids the entire doubling problem, but it introduces an
explicit chiral symmetry breaking effect on the quark mass. Such an effect can be suppressed by adding a clover term. The solution to avoid the entire fermion doubling problem without explicit chiral symmetry breaking is to use a Ginsparg-Wilson action~\cite{Ginsparg:1981bj}, such as the overlap fermion~\cite{Chiu:1998eu,Liu:2002qu}. More details of the fermion actions can be found in the Supplemental Materials~\cite{ref_sm}.

At the next-to-leading-order (NLO) of the PQ$\chi$PT, the mixed-action effect defined in Eq.~(\ref{eq:def_mix}), can be nonzero when $m_{\pi,{\rm vv}}\neq  m_{\pi,{\rm ss}}$, even if the fermion actions used by the valence and sea quarks are the same. 
Thus in this work, we majorly concentrate on the unitary case with the valence pion mass tuned to be the same as the sea pion mass,
\bal\label{eq:mix_action}
\Delta_{\rm mix, uni}(m_{\pi},a)\equiv \Delta_{\rm mix}(m_{\pi},m_{\pi},a).
\eal
$\Delta_{\rm mix, uni}$ approaches exactly zero in the continuum and thus is a good reference to verify the additional discretization error in the mixed fermion action simulation.  Such a definition is different from that used \lc{in} the previous $\chi$QCD study~\cite{Lujan:2012wg} but still provides consistent results, and detailed comparisons can be found in the Supplemental Materials~\cite{ref_sm}.

Since the HYP smearing~\cite{Hasenfratz:2001hp} can make the cost of the overlap fermion calculation to be much cheaper and the fluctuation of the clover fermion around the physical pion mass to be smaller, we will concentrate on the following three kinds of the valence fermion actions to calculate $\Delta_{\rm mix}$:

1) OV: Overlap fermion with 1-step HYP smearing and $\rho=1.5$;

2) HI: HISQ action without any additional smearing on the gauge link; 

3) CL: Clover fermion with 1-step HYP smearing and tree level tadpole improved clover coefficient $c_{sw}=\langle U_p\rangle ^{-3/4}$ where $\langle U_p\rangle$ is the vacuum expectation value of the plaquette on the HYP smeared configurations, calculated non-perturbatively on each gauge ensemble. 

\begin{table}[ht!]                   
\caption{Information of the ensembles~\cite{Fukaya:2010na,MILC:2010pul,MILC:2012znn,Bazavov:2017lyh,Blum:2014tka,Boyle:2015exm,Zhang:2021oja} used in this calculation.}  
\begin{tabular}{l l l l c l c }                                                                
\text{Action} &\text{Symbol} & $L^3 \times T$  &  $a$ (fm)   & $m_{\pi}$ (MeV)  \\
\hline   
DW+ID & 24D &$24^3\times\ 64$& 0.194& 139    \\  
DW+ID & 32Df &$32^3\times\ 64$& 0.143 & 143  &    \\    
DW+I & 48I &$48^3\times\ 96$& 0.114 & 139  \\ 
DW+I & 24I &$24^3\times\ 64$& 0.111 & 340   \\  
DW+I & 64I &$64^3\times128$& 0.084 & 139  \\          
DW+I & 32I &$32^3\times\ 64$& 0.083 & 302 \\   
\hline
OV$^{\rm 0HYP}$+IR & JLQCD &$24^3\times\ 48$&  0.112 & 290 \\                     
\hline
HI+S$^{(1)}$ & a12m310 &$24^3\times\ 64$& 0.121 & 310  \\  
HI+S$^{(1)}$  & a09m310 &$32^3\times\ 96$& 0.088 & 310   \\   
HI+S$^{(1)}$  & a06m310 &$48^3\times 144$& 0.057 & 310  \\  
HI+S$^{(1)}$ & a04m310 &$64^3\times 192$& 0.043 & 310   \\  
\hline
CL$^{\rm stout}$+S$^{\rm tad}$ & C11 &$24^3\times\ 72$& 0.108 & 290   \\   
CL$^{\rm stout}$+S$^{\rm tad}$ & C08 &$32^3\times\ 96$& 0.080 & 300    \\       
CL$^{\rm stout}$+S$^{\rm tad}$ & C06 &$48^3\times144$& 0.054 & 300   \\                     
\hline
\end{tabular}  
\label{tab:ensemble1}                                                                                  
\end{table}    

For the sea fermion actions, we use three kinds of the gauge ensembles to cover the popular choices: 

1) those from the RBC/UKQCD collaboration which use the 2+1 flavor DW fermion action and Iwasaki gauge action at four lattice spacings and physical pion mass~\cite{Blum:2014tka,Boyle:2015exm}, noting that the ensembles at the two largest lattice spacings include the DSDR term~\cite{Boyle:2015exm} (DW+ID, DD for short) and those at the finer two lattice spacings do not (DW+I, DW for short); 

2)  those from the MILC collaboration which use the 2+1+1 flavor HISQ fermion action with one-loop Symanzik improved gauge action~\cite{Hart:2008sq} at four lattice spacings and $m_{\pi}= 310$ MeV~\cite{MILC:2010pul,MILC:2012znn,Bazavov:2017lyh} (HI+S$^{(1)}$, HI for short); 

3) those from the CLQCD collaboration which use the 2+1 flavor tadpole improved clover fermion action (with 1-step stout link smearing on the gauge link with smearing parameter 0.125) and tadpole improved Symanzik gauge action at three lattice spacings and $m_{\pi}\simeq 300$ MeV~\cite{Zhang:2021oja} (CL$^{\rm stout}$+S$^{\rm tad}$, CL$^{\rm stout}$ for short).

We also repeat the calculation on a JLQCD ensemble~\cite{Fukaya:2010na} at $a=0.112$ fm and $m_{\pi,{\rm ss}}=290 \mathrm{MeV}$. This ensemble uses the 2+1 flavor overlap fermion ($\rho=1.3$ without any HYP smearing~\cite{Hasenfratz:2001hp} on the gauge link) and Iwasaki gauge  action with an additional ratio of extra Wilson fermions and associated twisted mass bosonic spinors~\cite{Fukaya:2006vs}. The brief information of all the ensembles used in this work are collected in Table~\ref{tab:ensemble1}.  More details of those ensembles can be found in the Supplemental Materials~\cite{ref_sm}.

On the DW/DD ensembles, we use the existing point source DW fermion propagators with the field sparsening compression~\cite{Li:2020hbj} and generate the valence quark propagators with similar source positions. On the other ensembles, we use the Coulomb gauge fixed wall source to take the advantage of the $L^3$ enhancement of statistics. Note that the wall source using the HISQ action is practically a grid source which picks only the even points in each spacial direction, as the Dirac space has been mapped to the even/odd sites and should be treated separately in the source. At the same time, we also use low-mode substitution~\cite{Li:2010pw} to suppress the statistical uncertainty of the pion correlators using the overlap fermion on the HISQ ensembles at small lattice spacings. The details of the unitary Chroma~\cite{Edwards:2004sx}+QUDA~\cite{Clark:2009wm,Babich:2011np,Clark:2016rdz} interface for various fermion actions can be found in Ref.~\cite{Zhang:2022tec}.

\begin{figure}[htbp]
    \centering
   \includegraphics[width=1\linewidth]{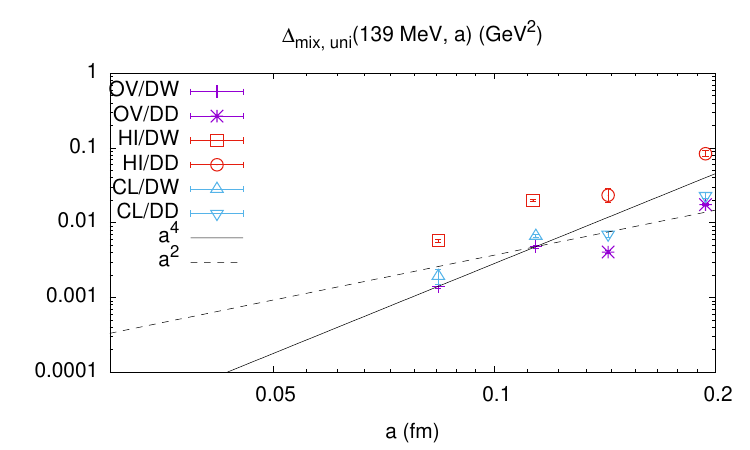}
    \caption{The mixed-action effect $\Delta_{\rm mix}$ on the DW and DD ensembles, as functions of the lattice spacing $a$. The symbol X/Y means the case with valence fermion action X on the sea fermion action Y. The figure also illustrates the $a^4$ (solid line) and $a^2$ (dashed line) dependence for comparison.}
    \label{fig:Sea_dw}
\end{figure}

{\it Results:} The lattice spacing $a$ dependence of $\Delta_{\rm mix, uni}$ on the DW/DD ensembles are shown in Fig.~\ref{fig:Sea_dw}, and the symbol X/Y corresponds to the case with valence fermion action X on the sea fermion action Y. If we take $\Delta_{\rm mix}^{\rm OV/DW}(a=0.114~\mathrm{fm})=0.00477(20)~\mathrm{GeV}^2$ and assume the lattice spacing dependence is $a^4$ (black line), then the prediction at 0.084 fm will be 0.00141(6) which agrees with the data 0.00141(9) perfectly. On the other hand, the prediction with a simple $a^2$ dependence (dashed line) will be 0.0026(1) which is more than 5$\sigma$ higher than the practical result. We can also solve $n$ from the data using the $C a^n$ form with different valence fermion actions; we obtain $n=$~4.0(2) (OV/DW, purple crosses), 4.0(2) (HI/DW, red boxes), and 4.0(7) (CL/DW, blue upward triangles), respectively. 

It is interesting to see that the similar calculations on two coarse lattice spacings 0.141 and 0.194 fm provide a similar power of the lattice spacing, as $n=$~4.8(2) (OV/DD, purple stars),  4.2(7) (HI/DD, red circles), and 3.9(4) (CL/DD, blue downward triangles), respectively; but they are a factor $\sim$2 smaller than $\Delta^{\rm X/DW}_{\rm mix}$ assuming an $a^4$ lattice spacing dependence, as illustrated by the black line for the $\Delta_{\rm mix}^{\rm OV/DW}$ case. Such a suppression of $\Delta_{\rm mix}$ would relate to the DSDR term, as the setup of the DD and DW ensembles are the same except this term. 

\begin{figure}[htbp]
    \centering
   \includegraphics[width=1\linewidth]{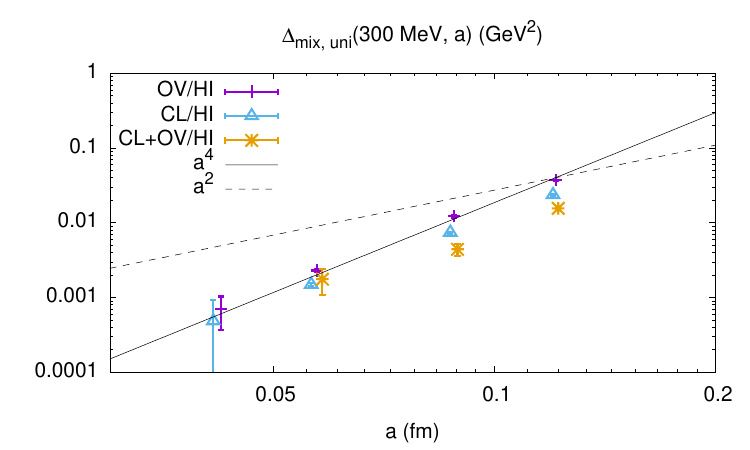}
    \caption{The mixed-action effects of the overlap fermion (purple crosses) or Clover fermion (blue upward triangles) on the HI+S$^{(1)}$ ensembles at four lattice spacings, together with that of the two valence fermion actions (yellow stars). The latter two types of data points are shifted horizontally to improve the visibility.}
    \label{fig:Sea_hisq}
\end{figure}

Based on the MAPQ$\chi$PT framework, the quark mass dependence of $\Delta_{\rm mix}$ is a next-to-leading order (NLO) correction and is thus relatively weak. It has been verified by our numerical calculations on both the DW+I and HI+S$^{(1)}$ ensembles~\cite{ref_sm}. Thus we turn to the HI+S$^{(1)}$ ensembles~\cite{MILC:2010pul,MILC:2012znn} with $\sim 310$ MeV pion mass to further investigate the lattice spacing dependence of $\Delta_{\rm mix}$ with the same simulation setup, and show the results in Fig.~\ref{fig:Sea_hisq}. 

With the data at four lattice spacings, it is more obvious that the $a^4$ behaviors observed on the DW/DD ensembles are not accidental and can dominate in a wide lattice spacing range. Note that the relative uncertainty at $a=0.043$ fm is large as the mixed action effect only changes the pion mass by $\sim$1(1) MeV there, and then one cannot exclude the possibility on the dominance of the lower order lattice spacing dependence (e.g. ${\cal O}(a^2)$ as expected by lattice perturbation theory~\cite{Bar:2003mh}). 

\begin{figure}[htbp]
    \centering
   \includegraphics[width=1\linewidth]{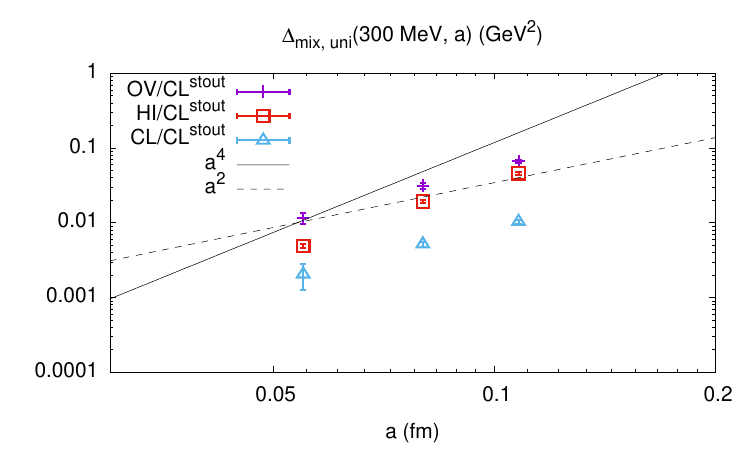}
    \caption{The mixed-action effects of the overlap fermion (OV, purple crosses), HISQ fermion (HI, red boxes) or HYP smeared clover fermion (CL, blue triangles) on the clover fermion ensembles at three lattice spacings,}
    \label{fig:Sea_clv}
\end{figure}

We also report the calculation on the CL$^{\rm stout}$+S$^{\rm tad}$ ensembles~\cite{Zhang:2021oja} in the lattice spacing range $a\in[0.054,0.108]$~fm. Unlike the DW+I or HI+S$^{(1)}$ ensembles using the chiral fermion, $\Delta_{\rm mix, uni}$ at $a=0.080$ and 0.108 fm look like an $a^2$ behavior, but the lattice spacing dependence becomes closer to $a^4$ from $a=0.080$ fm to 0.054 fm. Thus the behavior at larger lattice spacing could come from the cancellation between the $a^4$ and $a^6$ terms. At the same time, the mixed-action effect of the overlap fermion is larger than that of the HISQ fermion, which is also different from the behavior on the DW+I or HI+S$^{(1)}$  ensembles which use the chiral sea fermion actions. Since the CL$^{\rm stout}$+S$^{\rm tad}$ ensembles and HI+S$^{(1)}$ ensembles using similar Symanzik gauge actions with a minor difference in the 1-loop correction, the huge $\Delta_{\rm mix, uni}$ would be majorly related to the clover fermion action in the ensembles, while more accurate comparison with the exactly the same gauge action is worth further study in the future.

In order to further investigate the likeness between different fermion actions, we define the ``mixed-action effect'' of two valence fermion actions B and C on the same gauge ensemble with sea fermion action A as:
\bal\label{eq:def_mix2}
&\bar{\Delta}^{\rm B+C/A}_{\rm mix, uni}(m_{\pi},a)\equiv m_{\pi,{\rm BC}}^2\nonumber\\
&\quad \quad -\frac{m_{\pi,{\rm BB}}^2+m_{\pi,{\rm CC}}^2}{2}|_{m_{\pi,{\rm BB}}=m_{\pi,{\rm CC}}=m_{\pi,{\rm AA}}=m_{\pi}}.
\eal
As shown in Fig.~\ref{fig:Sea_hisq}, $\bar{\Delta}^{\rm CL+OV/HI}$ also decreases as $a^4$ while that at 0.06 fm has large uncertainty. At the same time, we will see that $\bar{\Delta}^{\rm B+C/A}_{\rm mix, uni}$ is very sensitive to the sea action $A$ which doesn't appear in the definition explicitly.

\begin{figure}[htbp]
    \centering
    \includegraphics[width=1\linewidth]{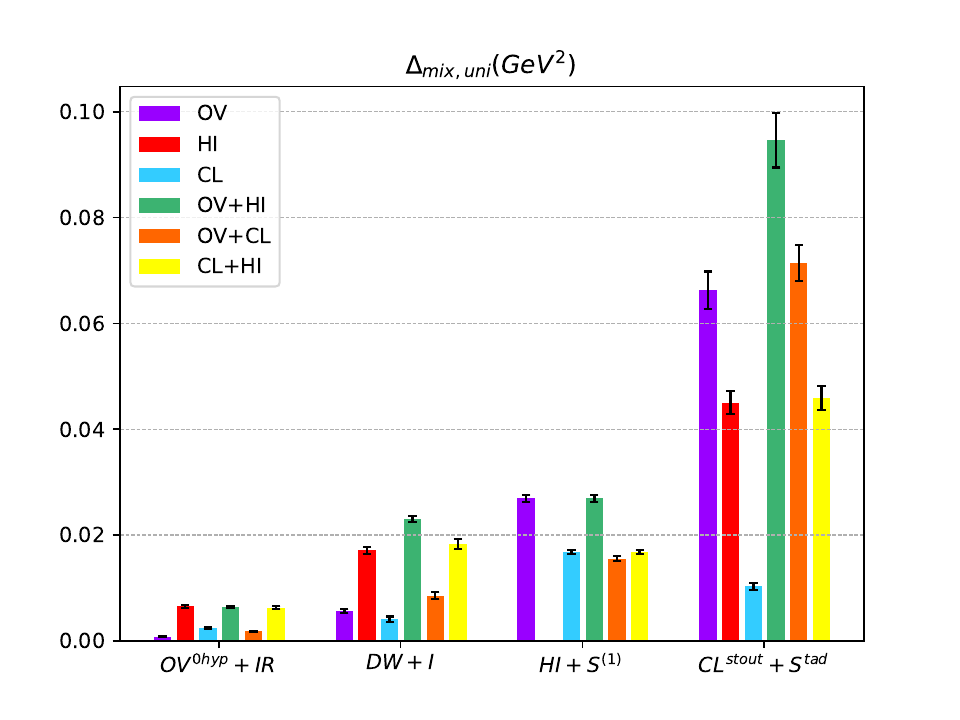}
    \caption{The mixed action effects $\Delta_{\rm mix, uni}$ for various valence fermion actions (OV, HI and CL) and $\bar{\Delta}_{\rm mix, uni}$ for different valence fermion actions (OV+HI, OV+CL and CL+HI) on the OV$^{\rm 0HYP}$+IR, DW+I, HI+S$^{(1)}$, and CL$^{\rm stout}$+S$^{\rm tad}$ ensembles. All the quantities are in unit of GeV$^2$, and are interpolated to $a=0.11$ fm and $m_{\pi}\in [290,310]$ MeV to make a fair comparison.}
    \label{fig:mix_all}
\end{figure}

To make a fair comparison of different cases, we do a linear $m_\pi^2$ interpolation based on MAPQ$\chi$PT, and also a linear $\log(a)$ interpolation on $\log \Delta_{\rm mix}(a)$ which is exact when $\Delta_{\rm mix}(a)\propto a^n$. Eventually, we show the interpolated values of $\Delta_{\rm mix, uni}$ and $\bar{\Delta}_{\rm mix, uni}$ on the OV$^{\rm 0HYP}$+IR, DW+I, HI+S$^{(1)}$ and CL$^{\rm stout}$+S$^{\rm tad}$ ensembles in Fig.~\ref{fig:mix_all}, for
$a\simeq 0.11$ fm and $m_{\pi}\in [290,310]$ MeV.

The red, blue and purple bars correspond to $\Delta_{\rm mix, uni}$ of the HISQ, clover and overlap valence fermion actions, respectively. Also, green, orange and yellow bars correspond to $\bar{\Delta}_{\rm mix, uni}$ of different valence fermion actions. The uncertainties of $\Delta_{\rm mix, uni}$ and $\bar{\Delta}_{\rm mix, uni}$ are also shown at the top of the bar in the figure. In addition, the HISQ valence quark is the same as the sea quark under the HI+S$^{(1)}$ ensemble, therefore the `HI' bar in the third group of this figure is exactly zero.

From the figure, the mixed-action effect on the OV$^{\rm 0HYP}$+IR ensemble seems to be smaller than the DW+I case, which is comparable with the DW+ID case given the factor of 2 suppression shown in Fig.~\ref{fig:Sea_dw}. It is predictable as the DSDR term provides somehow similar impact on the near-zero modes of the Dirac operator as the ratio term~\cite{Fukaya:2006vs,Vranas:2006zk}. 

Compared to the mixed-action effects on the DW+I ensembles, those on the HI+S$^{(1)}$ ensemble are somehow similar. But when we consider the valence CL or OV action, the $\Delta_{\rm mix, uni}$ on the HI+S$^{(1)}$ ensemble are still larger than those on the DW+I ensemble. At the same time, the impact of the gauge action used by DW+I or HI+S$^{(1)}$ is minor, which requires further study using different fermion actions and the same gauge action.

Contrary to the above three cases, the mixed-action effects on the CL$^{\rm stout}$+S$^{\rm tad}$ ensemble are much larger, even for the case of the HYP smeared clover valence fermion action on stout smeared clover sea action. The origin of this huge mixed-action effect presumably relates to the additive chiral symmetry breaking of the clover fermion action.

Besides the sensitivity on the sea actions, our results also suggest that $\bar{\Delta}_{\rm mix, uni}$ satisfies the following triangle inequalities within the statistical uncertainty,
\bal\label{eq:ineq}
|\Delta^{\rm B/A}_{\rm mix, uni}-\Delta^{\rm C/A}_{\rm mix, uni}|&\le \bar{\Delta}^{\rm B+C/A}_{\rm mix, uni} \le \Delta^{\rm B/A}_{\rm mix, uni}+\Delta^{\rm C/A}_{\rm mix, uni},\\
|\bar{\Delta}^{\rm B+D/A}_{\rm mix, uni}-\bar{\Delta}^{\rm C+D/A}_{\rm mix, uni}|&\le \bar{\Delta}^{\rm B+C/A}_{\rm mix, uni} \le \bar{\Delta}^{\rm B+D/A}_{\rm mix, uni}+\bar{\Delta}^{\rm C+D/A}_{\rm mix, uni}.
\eal
where D is a valence fermion action different from B and C. The inequalities in Eq.~(\ref{eq:ineq}) would have fundamental origin other than simple empirical observations, as both the sea action sensitivity and the $a^4$ behavior of $\bar{\Delta}_{\rm mix, uni}$ can be deduced from Eq.~(\ref{eq:ineq_1}). But MAPQ$\chi$PT just includes one kind of valence fermion action and should be extended to describe our observations here. The numerical results of $\Delta_{\rm mix, uni}$ and $\bar{\Delta}_{\rm mix, uni}$ in all the cases we studied, and also the illustration of Fig.~\ref{fig:mix_all} as triangles, can be found in the Supplemental Material~\cite{ref_sm}. 

{\it Summary}: Based on the calculation of the mixed-action pion mass using three kinds of the fermion actions, including {overlap} fermion, HISQ fermion, and clover Wilson fermion, on different kinds of the dynamical ensembles, we found that the leading mixed-action effect
is numerically small and decreases with small lattice spacing as $a^4$.
It is particularly small when the sea fermion action has chiral symmetry, much smaller than the naive ${\cal O}(a^2)$ estimate at small lattice spacings. We note that a similar $a^4$ behavior was observed in the taste mixing effect of the HISQ fermion in Ref.~\cite{Borsanyi:2020mff,Aubin:2022hgm}. 

Based on MAPQ$\chi$PT, $\Delta_{\rm mix, uni}$ is the only leading order low energy constant so that we expect the mixed-action effect in other quantities will also be ${\cal O}(a^4)$, even though this expectation should be verified by the practical calculations. Thus our study provides the first quantitative estimate on the systematic uncertainty from the mixed-action lattice QCD simulation, and potentially releases the restriction to do the simulation using the gauge ensemble with cheaper fermion actions but larger volume, smaller lattice spacing and light pion mass. More specifically, an ensemble using a fermion action satisfying the Ginsburg-Wilson relation and Iwasaki gauge action can provide the smallest mixed-action effect, and the ratio or DSDR term can further suppress it with the price of the topological charge fluctuation suppression~\cite{Fukaya:2010na,Boyle:2015exm}. The HISQ fermion and one-loop Symanzik improved gauge action ensembles can be an acceptable choice for all the other more expensive fermion actions at small lattice spacing thanks to the $a^4$ behavior. On the other hand, the ensembles using the clover fermion action have much larger mixed-action effects for all the three kinds of the valence fermion actions, and thus the mixed-action calculation on those ensembles can be more challenging. The mixed-action effects studied in this work can also be served as one benchmark for the differences of results obtained with different fermion actions.

\section*{Acknowledgment}
We thank the CLQCD, JLQCD, MILC and RBC/UKQCD collaborations for providing us their gauge configurations with dynamical fermions, and Terrence Draper for valuable discussion and comments. The calculations were performed using the Chroma software suite~\cite{Edwards:2004sx} with QUDA~\cite{Clark:2009wm,Babich:2011np,Clark:2016rdz}, GWU code~\cite{Alexandru:2011ee,Alexandru:2011sc} and OpenChiral through HIP programming model~\cite{Bi:2020wpt}.  The domain wall propagators were generated using Grid, GPT, and Qlattice.
The numerical calculations in this study were carried out on the ORISE Supercomputer and HPC Cluster of ITP-CAS. This research used resources of the Oak Ridge Leadership Computing Facility at the Oak Ridge National Laboratory, which is supported by the Office of Science of the U.S. Department of Energy under Contract No.\ DE-AC05-00OR22725. This work used Stampede time under the Extreme Science and Engineering Discovery Environment (XSEDE), which is supported by National Science Foundation Grant No.\ ACI-1053575.
We also thank the National Energy Research Scientific Computing Center (NERSC) for providing HPC resources that have contributed to the research results reported within this paper.
We acknowledge the facilities of the USQCD Collaboration used for this research in part, which are funded by the Office of Science of the U.S. Department of Energy. Y. Yang is supported in part by the Strategic Priority Research Program of Chinese Academy of Sciences, Grant No.\ XDB34030303 and XDPB15, and also the National Natural Science Foundation of China (NSFC) under Grants No. 12047503.. P. Sun is supported in part by the National Natural Science Foundation of China under Grant No.\ 11975127. Y. Yang and P. Sun are also supported in part by a NSFC-DFG joint grant under grant No. 12061131006 and SCHA 458/22 and the GHfund A No.\ 202107011598. G. Wang is supported by the U.S. DOE Grant No.\ DE-SC0013065, DOE Grant No.\ DE-AC05-06OR23177 which is within the framework of the TMD Topical Collaboration and the French National Research Agency under the contract ANR-20-CE31-0016. L.C.J. acknowledges support by DOE Office of Science Early Career Award DE-SC0021147 and DOE grant DE-SC0010339.

\bibliographystyle{apsrev4-1}
\bibliography{reference.bib} 

\clearpage
%
%
%
%
%
%
%

\begin{widetext}

\section*{Supplemental materials}

\subsection{Note on the discretized fermion and gauge actions}

The naive fermion action below has the infamous fermion doubling problem. In 4 dimensions, it leads to 16 fermions as opposed to 1,
\bal
&S_{\rm naive}(m)=\sum_{x,y}\bar{\psi}(x)D_{\rm naive}(x,y,m)\psi(y),
\eal
with $D_{\rm naive}(x,y,m)=\frac{1}{2}\sum_{\mu=1,...,4,\eta=\pm}\big[\eta\gamma_{\mu}U_{\mu}(x,x+\eta\hat{n}_{\mu}a)\delta_{y,x+\eta\hat{n}_{\mu}a}\big] -ma\delta_{y,x}$. The Staggered fermion 
\bal
&S_{\rm st}(m)=\sum_{x,y}\bar{\psi}_{\rm st}(x)D_{\rm st}(x,y,m)\psi_{\rm st}(y),
\eal
introduces a redefinition of the quark field (e.g., $\psi_{\rm st}(x)=\gamma_4^{x_4}\gamma_1^{x_1} \gamma_2^{x_2} \gamma_3^{x_3}\psi(x)$ at the site $x=\{x_1,x_2,x_3,x_4\}$ in the MILC convention) and rotates $D_{\rm naive}$ into 
$D_{\rm st}(x,y,m)=\frac{1}{2}\sum_{\mu=1,...,4,\eta=\pm}\big[\eta \eta_{\mu} U_{\mu}(x,x+\eta\hat{n}_{\mu}a) \delta_{y,x+\eta\hat{n}_{\mu}a}\big] -ma\delta_{y,x}$, where $\{\eta_{1},\eta_2,\eta_3,\eta_4\}=\{(-1)^{x_4},(-1)^{x_1+x_4},(-1)^{x_1+x_2+x_4},1\}$
in the MILC convention.
Staggered fermions do not completely fixed the fermion doubling problem, as
there are 4 residual fermions, usually named as ``tastes'' of fermion.
Further improvement such as using the HISQ action~\cite{Follana:2006rc} is needed to suppress the mixing between the tastes. 

On the other hand, the Wilson fermion
\bal
&S_{\rm w}(m)=\sum_{x,y}\bar{\psi}(x)D_{\rm w}(x,y,m)\psi(y)
\eal
with $D_{\rm w}(x,y,m)=D_{\rm naive}(x,y,m)+\frac{1}{2}\sum_{\mu=1,...,4,\eta=\pm}U_{\mu}(x,x+\eta\hat{n}_{\mu}a)\delta_{y,x+\eta\hat{n}_{\mu}a} -4\delta_{y,x}$
avoids the entire doubling problem. But it introduces an
explicit chiral symmetry breaking effect on the quark mass. This can be improved by adding a clover term:
 \bal
 &S_{\rm cl}(m)=S_{\rm w}(m)+c_{\rm sw} \sum_{x}\bar{\psi}(x)\sigma_{\mu\nu}F^{\mu\nu}\psi(x),
 \eal
 where $c_{\rm sw}=1+{\cal O}(\alpha_s)$ is a tunable parameter.

The solution to avoid the fermion doubling problem entirely without explicit chiral symmetry breaking is to use a Ginsparg-Wilson action~\cite{Ginsparg:1981bj}, such as the overlap fermion~\cite{Chiu:1998eu,Liu:2002qu},
\bal
S_{\rm c}(m)=\sum_x \bar{\psi}(x)[(1-\frac{m}{2\rho})D_{\rm ov}(x,y,m)+\delta_{x,y}m]\psi(y),
\eal
where $D_{\rm ov}=\rho(1+\frac{D_{\rm w}(-\rho)}{\sqrt{D^{\dagger}_{\rm w}(-\rho)D_{\rm w}(-\rho)}})$ with a positive $\rho$ such as $\rho=1.5$.

The gauge actions in the ensembles we used in this work can be considered as particular cases of the improved L\"uscher-Weisz action~\cite{Luscher:1984xn},
\bal
S=\sum_x \mathrm{Re}\big[ c_0U_{\rm plaquettes}+c_1U_{\rm Rectangles}+c_2U_{\rm parallelograms}\big],
\eal
where $c_{0,1,2}$ are coefficients to be determined perturbatively or non-perturbatively, $U_{\rm plaquettes}$ and $U_{\rm Rectangles}$ are the standard $1\times 1$ and $1\times 2$ Wilson loops, and $U_{\rm parallelograms}$ are the ``parallelogram'' Wilson loops with 6 links defined in Ref.~\cite{Luscher:1984xn}. 

\begin{table}[ht!]                   
\caption{Information of the ensembles~\cite{MILC:2010pul,MILC:2012znn,Bazavov:2017lyh,Blum:2014tka,Boyle:2015exm,Zhang:2021oja,Fukaya:2010na} used in this work. The symbol I corresponds to the standard Iwasaki action, ID/IR means the Iwasaki action with extra DSDR or JLQCD ratio terms used in the simulation, and S$^{(1)}$/S$^{\rm tad}$ is used for the Symanzik gauge action with fully 1-loop improvement or simple tree-level tadpole improvement, respectively.}\label{tab:gauge_coeff}
\begin{tabular}{l l l | c c c  c | c  c  }                                                                
\text{Action} &\text{Symbol} & $L^3 \times T$ & $c_0$ & $c_1$ & $c_2$  & $u_0$  &  $a$ (fm)   & $m_{\pi,{\rm ss}}$ (MeV)  \\
\hline   
DW+ID & 24D &$24^3\times\ 64$& 5.957& -0.541 & 0 & 0.827 & 0.194& 139   \\  
DW+ID & 32Df &$32^3\times\ 64$& 6.384 & -0.579 & 0 & 0.846 & 0.143 & 143    \\    
\hline   
DW+I & 48I &$48^3\times\ 96$& 7.770 & -0.705 & 0 & 0.875 & 0.114 & 139  \\ 
DW+I & 24I &$24^3\times\ 64$& 7.770 & -0.705 & 0 & 0.876 & 0.111  & 340  \\  
DW+I & 64I &$64^3\times128$ & 8.208 & -0.745 & 0 & 0.886 & 0.084 & 139  \\        
DW+I & 32I &$32^3\times\ 64$& 8.208 & -0.745 & 0 & 0.886 & 0.083  & 302   \\           
\hline
OV$^{\rm 0HYP}$+IR & JLQCD &$24^3\times\ 48$& 8.390 & -0.761 & 0 & 0.884 & 0.112 & 290 \\           
\hline
HI+S$^{(1)}$ & a12m310 &$24^3\times\ 64$& 6.000 & -0.298 & 0.022 & 0.864 & 0.121 & 310  \\  
HI+S$^{(1)}$ & a12m130 &$48^3\times\ 64$& 6.000 & -0.298 & 0.022 & 0.864 & 0.121 & 130  \\  
HI+S$^{(1)}$  & a09m310 &$32^3\times\ 96$& 6.300 & -0.281 & 0.021 & 0.874 & 0.088 & 310   \\   
HI+S$^{(1)}$  & a09m130 &$64^3\times\ 96$& 6.300 & -0.281 & 0.021 & 0.874 & 0.088 & 130   \\   
HI+S$^{(1)}$  & a06m310 &$48^3\times 144$& 6.720 & -0.263 & 0.020 & 0.886 & 0.057 & 310  \\  
HI+S$^{(1)}$ & a04m310 &$64^3\times 192$& 7.000 & -0.254 & 0.019 & 0.892 & 0.043 & 310   \\  
\hline
CL$^{\rm stout}$+S$^{\rm tad}$ & C11 &$24^3\times\ 72$& 6.200 & -0.440 & 0 & 0.855 & 0.108 & 290   \\   
CL$^{\rm stout}$+S$^{\rm tad}$ & C08 &$32^3\times\ 96$& 6.410 & -0.430 & 0 & 0.863 & 0.080 & 300    \\       
CL$^{\rm stout}$+S$^{\rm tad}$ & C06 &$48^3\times144$& 6.720 & -0.424 & 0 & 0.873 & 0.055 & 300   \\
\hline
\end{tabular}  
\end{table}

On the DW+I~\cite{Blum:2014tka,Boyle:2015exm} and ~\cite{Fukaya:2010na} ensembles, the Iwasaki gauge action uses the fixed coefficients $c_{0,1,2}$ at arbitrary lattice spacing,
\bal
c_0=(1-8 c) \beta,\ c_1=c\beta,\ c_2=0,
\eal
where $c=-0.331$, $\beta=6/g^2$ and $g$ is the bare coupling constant. On the OV$^{\rm 0HYP}$+IR ensemble, extra Wilson fermions and associated twisted mass bosonic spinors are introduced to generate a weight
\bal
{\rm det}[H_W(-\rho)]/{\rm det}[H_W^2(-\rho)+m_t^2]
\eal
with $m_t\sim 0.2$ in the functional integrals, to suppress the near-zero mode of $H_W(-\rho)$~\cite{Fukaya:2006vs}. Similarly, the DW+ID ensemble uses an improved version of the above weight as the dislocation suppressing determinant ratio (DSDR)~\cite{Vranas:2006zk} term,
\bal
{\rm det}[H_W^2 (-\rho)+\epsilon_f^2]/{\rm det}[H_W^2 (-\rho)+\epsilon_b^2]
\eal
with $\epsilon_f=0.02$ and $\epsilon_b=0.5$. As shown in Table~\ref{tab:gauge_coeff}, the extra weight can make the $c_0$ of the OV$^{\rm 0HYP}$+IR ensemble at 0.11 fm to be $\sim 8$\% larger than that of the DW+I ensemble at similar lattice spacing. 

On the other hand, the one-loop improved Symanzik gauge action in the HI+S$^{(1)}$ ensembles~\cite{MILC:2010pul,MILC:2012znn} uses the lattice spacing dependent coefficients~\cite{Hart:2008sq} with $N_f=4$,
\bal
c_0=\beta',\ c_1=-\frac{\beta'}{20 u_0^2} [1-(0.6264-1.1746 N_f)]\mathrm{ln} u_0,\ c_2= \frac{\beta'}{u_0^2} (0.0433-0.0156N_f) \mathrm{ln} u_0,
\eal
where $u_0=\langle \frac{1}{18}U_{\rm plaquattes}\rangle^{1/4}$ is the tadpole improvement factor which can be determined self-consistently during the HMC, and the $\mathrm{ln} u_0$  dependence reflects the 1-loop correction through the approximation $\alpha_s\simeq -\frac{4 \mathrm{ln} u_0}{3.0684}$~\cite{Orginos:1998ue}. Note that $\beta'=\frac{5}{3u_0^4} \beta$ also includes the additional factors in front of the plaquette term and then differs from $\beta$ by a factor $\sim 3$.

At the same time, the gauge action used by the CL$^{\rm stout}$+S$^{\rm tad}$ ensembles~\cite{Zhang:2021oja} is the tree level Symanzik gauge action without the $\mathrm{ln} u_0$ terms,
\bal
c_0=\beta',\ c_1=-\frac{\beta'}{20 u_0^2} ,\ c_2= 0.
\eal
Based on the information shown in Table~\ref{tab:gauge_coeff}, the similar Symanzik gauge actions used by the HI+S$^{(1)}$ and CL$^{\rm stout}$+S$^{\rm tad}$ ensembles with the same $c_0$ provide approximately equal lattice spacing, even though they use different sea fermion actions and different $c_{1,2}$. But the value of $u_0$ seems to be quite sensitive to those differences.

\subsection{Comparison with the previous $\chi$QCD $\Delta_{\rm  mix}$ calculation}

At the next-to-leading-order (NLO) of partially quenched $\chi$PT, the pion masses used in the mixed action effect definition 
\bal\label{eq:def_mix_sm}
\Delta_{\rm mix}(m_{\pi,{\rm vv}},m_{\pi,{\rm ss}},a)&\equiv m_{\pi,{\rm vs}}^2-\frac{m_{\pi,{\rm vv}}^2+m_{\pi,{\rm ss}}^2}{2},
\eal
can be expressed as~\cite{Sharpe:1997by},
\bal
m^2_{\pi,{\rm vv}}&=\Lambda_{\chi}^2  2 y_{\rm v}\big\{1+\frac{2}{N_f}[(2 y_{\rm v}-y_{\rm s})\mathrm{ln} (2y_{\rm v})+(y_{\rm v}-y_{\rm s})]+2y_{\rm v} (2\alpha_8-\alpha_5)+2y_{\rm s} N_f (2\alpha_6-\alpha_4)+{\cal  O}(y^2)\big\},\\
m^2_{\pi,{\rm vs}}&=\Lambda_{\chi}^2  (y_{\rm v}+y_{\rm s})\big\{1+\frac{2}{N_f}y_{\rm v} \mathrm{ln} (2y_{\rm v})+(y_{\rm v}+y_{\rm s}) (2\alpha_8-\alpha_5)+2y_{\rm s} N_f (2\alpha_6-\alpha_4)+{\cal  O}(y^2)\big\},\\
m^2_{\pi,{\rm ss}}&=\Lambda_{\chi}^2  2 y_{\rm s}\big\{1+\frac{2}{N_f}y_{\rm s}\mathrm{ln} (2y_{\rm s})+2y_{\rm s} [2\alpha_8-\alpha_5+N_f (2\alpha_6-\alpha_4)]+{\cal  O}(y^2)\big\},
\eal
where $\Lambda_{\chi}=4\pi F$ is the intrinsic scale of $\chi$PT with $F$ being the pion decay constant in the chiral limit, $y=\frac{\Sigma m}{F^2 \Lambda_{\chi}^2}$ is the dimensionless 
expansion parameter with $\Sigma$ and $m$ being the chiral condensate and quark mass respectively, and $\alpha_i=128 \pi^2 L_i$ is the NLO low  energy constants of $\chi$PT.  Thus $\Delta_{\rm mix}(m_{\pi,{\rm vv}},m_{\pi,{\rm ss}},a)$ can be non-zero when $m_{\pi,{\rm vv}}\neq  m_{\pi,{\rm ss}}$, even the fermion actions used by the valence and sea quarks are the same. 

\begin{figure}[htbp]
    \centering
    \includegraphics[width=0.5\linewidth]{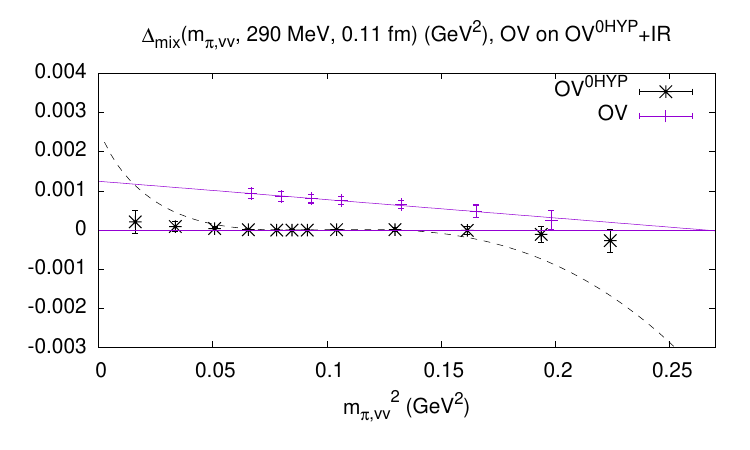}
    \caption{$\Delta_{\rm mix}(m_{\pi,{\rm vv}},290~\mathrm{MeV}, 0.11~\mathrm{fm})$ on the OV$^{\rm 0HYP}$+IR ensemble. The $\Delta_{\rm mix}$ using exactly the same action for both the valence and sea quarks (black stars) is consistent with zero, while using the overlap fermion action with HYP smeared link and $\rho=1.5$ (purple crosses) can make $\Delta_{\rm mix}$ to be obviously nonzero while still small. The NLO PQ$\chi$PT prediction of $\Delta_{\rm mix}$ is also shown as a black dashed line for comparison.}
    \label{fig:JLQCD}
\end{figure}

To quantify this effect, 
we carried out an unitary-action calculation on the OV$^{\rm 0HYP}$+IR ensemble~\cite{Fukaya:2010na} from the JLQCD collaboration at $a=0.112$ fm and $m_{\pi,{\rm ss}}=290~\mathrm{MeV}$. In Fig.~\ref{fig:JLQCD}, the black stars show the partially quenched (PQ) results with the same OV$^{\rm 0HYP}$ fermion action as that used in the ensemble, and the black dashed curve shows the NLO $\chi$PT prediction of $\Delta_{\rm mix}(m_{\pi,{\rm vv}},m_{\pi,{\rm ss}}=290~\mathrm{MeV})$ using the low-energy constants from the literature. They agree with each other well when $m_{\pi,{\rm vv}}$ is close to $m_{\pi,{\rm ss}}$, and it is understandable that the lattice results can be somehow different from the $\chi$PT prediction due to the finite volume effect when $m_{\pi,{\rm vv}}$ is small (or due to higher order $\chi$PT corrections when $m_{\pi,{\rm vv}}>400~\mathrm{MeV}$).

We also calculate $\Delta_{\rm mix}$ using the overlap fermion with HYP smeared gauge link and $\rho=1.5$ which is the default $\chi$QCD setup, and show the result (purple crosses) in Fig.~\ref{fig:JLQCD}. In such a case, the mixed-action effect at the unitary pion mass $\Delta_{\rm mix, uni}(m_{\pi,{\rm ss}})\equiv \Delta_{\rm mix}(m_{\pi,{\rm ss}},m_{\pi,{\rm ss}})=0.0008(1)$ is nonzero, and depends on the valence pion mass obviously. 

In this work, we majorly concentrate on the unitary case with the valence pion mass tuned to be the same as the sea pion mass,
\bal\label{eq:mix_action_sm}
\Delta_{\rm mix, uni}(m_{\pi},a)\equiv \Delta_{\rm mix}(m_{\pi},m_{\pi},a).
\eal
$\Delta_{\rm mix, uni}$ approaches exactly zero in the continuum and thus provides a good reference to check the additional discretization error in the mixed fermion action simulations.  Such a definition is different from that used the previous $\chi$QCD study~\cite{Lujan:2012wg}, in which the mixed-action effect is defined by the following parametrization,
\bal\label{eq:mix_action_old}
m_{vs}^2-\frac{1}{2}m^2_{ss}=B m_v +\tilde{\Delta}_{\rm mix},
\eal
and calculated on two DW+I ensembles with $\sim$300 MeV pion mass~\cite{Blum:2014tka}. Based on above definition, Ref.~\cite{Lujan:2012wg} gave $\tilde{\Delta}^{\rm OV/DW}_{\rm mix}(0.083~\mathrm{fm})=0.004(2)$  and $\tilde{\Delta}^{\rm OV/DW}_{\rm mix}(0.111~\mathrm{fm})=0.009(2)~\mathrm{GeV}^2$ (based on Table V of Ref.~\cite{Lujan:2012wg} with Method II). If we assume a simple $C a^n$ form and solve $n$, the data suggests an $a^{2.4(1.2)}$ behavior with large uncertainty on the power. 

\begin{figure}[htbp]
    \centering
   \includegraphics[width=0.5\linewidth]{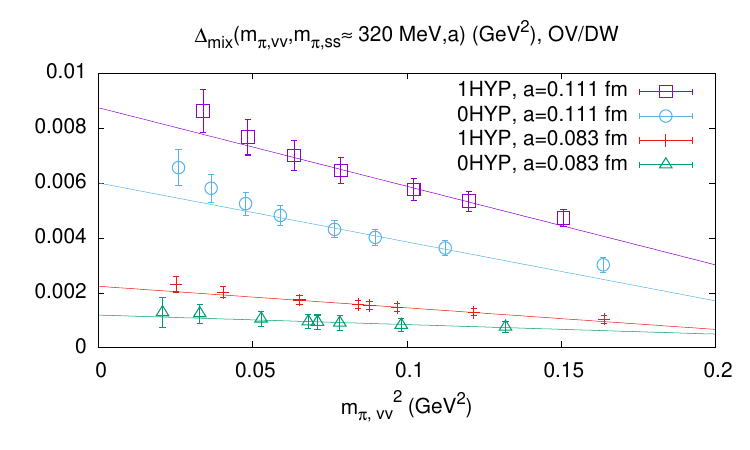}
    \caption{The mixed-action effect of the overlap fermion on the RBC/UKQCD Domain wall fermion ensembles~\cite{Blum:2014tka}, as functions of the valence pion mass $m_{\pi, \rm{vv}}$ with two lattice spacings and with or without HYP smearing. The lines show the fits in  the range of $m_{\pi, \rm{vv}}\in[0.05,0.15]~\mathrm{GeV}^2$.}
    \label{fig:DW+I}
\end{figure}

We repeated the calculation on 24I and 32I ensembles with several valence quark masses, and show the result as functions of valence pion mass in Fig.~\ref{fig:DW+I}. Based on the leading order $\chi$PT, $\tilde{\Delta}_{\rm mix}$ corresponds to the $\Delta_{\rm mix}$ in the chiral limit of $m_{\pi, \rm{vv}}$. As shown in the figures, the negative $m_{\pi, \rm{vv}}$ dependence of $\Delta_{\rm mix}$ (purple boxes for 24I and red cross for 32I) makes the $\tilde{\Delta}_{\rm mix}$ (illustrated as the intercepts at $m_{\pi, \rm{vv}}=0$ at the leading order of PQ$\chi$PT) to be larger than $\Delta_{\rm mix, uni}$, and consistent with the previous study within the uncertainty. But with much more precise data, similar parametrization suggests an $a^{4.6(0.4)}$ behavior which means that the mixed-action effect at small lattice spacings can be significantly smaller than that predicted by a naive $a^2$ correction. 

The results of $\Delta_{\rm mix}$ using the overlap valence fermion without HYP smearing are also presented in Fig.~\ref{fig:DW+I} as the blue dots (24I) and green triangles (32I). As shown in the figure, $\Delta_{\rm mix}$ without HYP smearing is $\sim$ 20--30\% smaller than the HYP smeared cases at all the $m_{\pi, \rm{vv}}$, while it has a similar $a^4$ lattice spacing dependence. 

\subsection{Quark mass dependence}

In the main text, we show the lattice spacing dependence of $\Delta_{\rm mix, uni}$ on the DW+I ensemble at physical pion mass, and also those on HI+S$^{(1)}$ at $m_{\pi}\simeq$ 300 MeV. In Fig.~\ref{fig:Sea_dw_mass} and Fig.~\ref{fig:Sea_hisq_mass}, we show the $\Delta_{\rm mix, uni}$ on those two ensembles  with both the physical pion mass (left panels) and $m_{\pi}\simeq$ 300 MeV (right panels). Even though the value of $\Delta_{\rm mix, uni}$ can depend on the pion mass at given lattice spacing, the lattice spacing dependence is always $a^4$ in all the cases.

\begin{figure}[H]
    \centering
   \includegraphics[width=0.45\linewidth]{figures/Sea_dw.pdf}
   \includegraphics[width=0.45\linewidth]{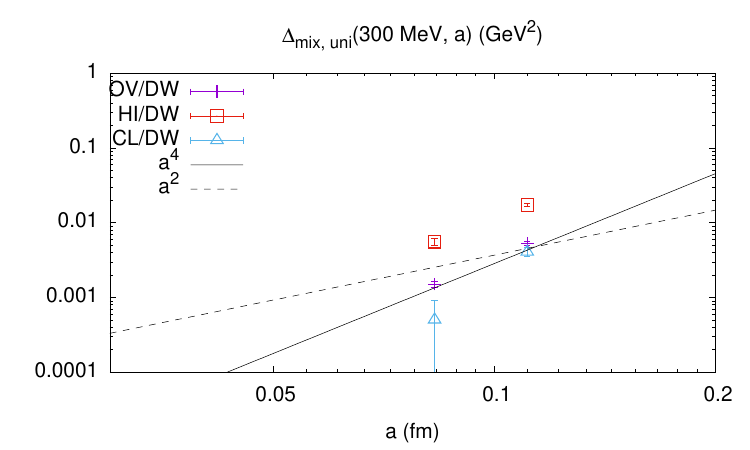}
    \caption{The mixed action effect $\Delta_{\rm mix, uni}$ on the DW+I ensembles, as functions of the lattice spacing $a$. The left panel shows the physical pion mass case and the right panel shows that using 300 MeV pion mass. The figures also illustrate the $a^4$ (solid line)  and $a^2$ (dash line) dependence for comparison.}
    \label{fig:Sea_dw_mass}
\end{figure}

\begin{figure}[H]
    \centering
    \includegraphics[width=0.45\linewidth]{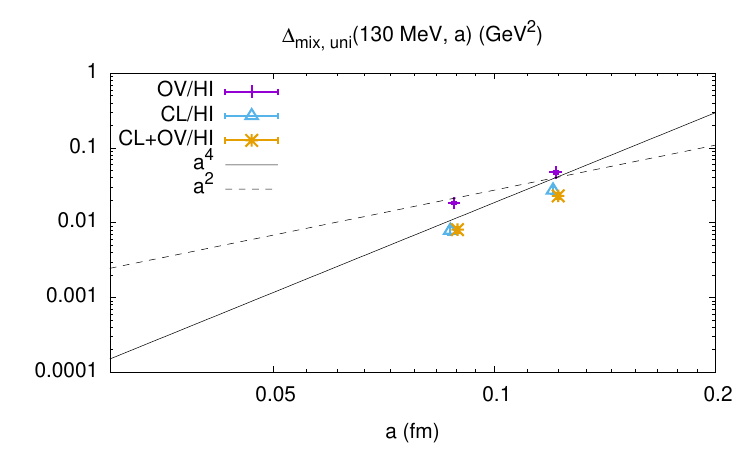}
   \includegraphics[width=0.45\linewidth]{figures/Sea_hisq.pdf}
    \caption{The mixed-action effects of the overlap fermion (purple crosses) or clover fermion (blue boxes) on the HI+S$^{(1)}$ ensembles using the HISQ fermion at four lattice spacings, together with that of the two valence fermion actions (yellow stars). The left panel shows the physical pion mass case and the right panel shows that using a 300 MeV pion mass. The latter two types of the data points are shifted horizontally to improve the visibility.}
    \label{fig:Sea_hisq_mass}
\end{figure}

\subsection{Numerical results of mixed-action effects under different ensembles}

In order to investigate the likeness between different fermion actions, we define the ``mixed-action effect'' of two valence fermion actions B and C on the same gauge ensemble with sea fermion action A as:
\bal
&\bar{\Delta}^{\rm B+C/A}_{\rm mix, uni}(m_{\pi},a)\equiv m_{\pi,{\rm BC}}^2\nonumber\\
&\quad \quad -\frac{m_{\pi,{\rm BB}}^2+m_{\pi,{\rm CC}}^2}{2}|_{m_{\pi,{\rm BB}}=m_{\pi,{\rm CC}}=m_{\pi,{\rm AA}}=m_{\pi}}.
\eal
As shown in Fig.~\ref{fig:Sea_hisq_mass}, $\bar{\Delta}^{\rm CL+OV/HI}$ also decreases as $a^4$ while that at 0.06 fm has large uncertainty, and satisfies the following inequalities within statistical uncertainty,
\bal\label{eq:ineq_1}
|\Delta^{\rm B/A}_{\rm mix, uni}-\Delta^{\rm C/A}_{\rm mix, uni}|&\le \bar{\Delta}^{\rm B+C/A}_{\rm mix, uni} \le \Delta^{\rm B/A}_{\rm mix, uni}+\Delta^{\rm C/A}_{\rm mix, uni}.
\eal
Furthermore, based on similar calculationa on the RBC/UKQCD ensemble\lc{s} for three kinds of valence fermion actions, the following relations are also satisfied for $\bar{\Delta}^{\rm B+C/A}_{\rm mix, uni}$,
\bal\label{eq:ineq_2}
|\bar{\Delta}^{\rm B+D/A}_{\rm mix, uni}-\bar{\Delta}^{\rm C+D/A}_{\rm mix, uni}|&\le \bar{\Delta}^{\rm B+C/A}_{\rm mix, uni} \le \bar{\Delta}^{\rm B+D/A}_{\rm mix, uni}+\bar{\Delta}^{\rm C+D/A}_{\rm mix, uni},
\eal
where D is a valence fermion action different from B and C.

In Table \ref{tab:mix_act}, we list all calculated $\Delta_{\rm mix, uni}$ and also $\bar{\Delta}_{\rm mix, uni}$ under different ensembles. The first four columns provide information of the ensembles similar to that shown in Table \ref{tab:ensemble1}, and the remaining six columns show $\Delta_{\rm mix, uni}$ and $\bar{\Delta}_{\rm mix, uni}$ on the corresponding ensembles with different valence quark actions.

\begin{table}[ht!]                   
\caption{Mixed-action effects on different ensembles with different valence quark actions.}\label{tab:mix_act}
\begin{tabular}{c c c c| c c c c c c}                                      
\multicolumn{4}{c}{} & \multicolumn{3}{c}{$\Delta_{\rm mix, uni}$(\text{GeV}$^2$)} & \multicolumn{3}{c}{$\bar{\Delta}_{\rm mix, uni}$(\text{GeV}$^2$)} \\
\text{Action} &\text{Symbol} &\text{a(fm)} &$m_{\pi,{\rm ss}}$ (MeV) &\text{OV} &\text{HI} &\text{CL} &\text{OV+HI} &\text{OV+CL} &\text{CL+HI}   \\
\hline   
DW+ID & 24D&0.194&139 &0.0174(04) & 0.0836(77) & 0.0230(21)& 0.0858(43) & 0.0345(23) & 0.0847(54)\\  
DW+ID & 32Df&0.143&143 &0.0040(02) & 0.0233(49) & 0.0070(07)& 0.0248(27) & 0.0097(11) & 0.0229(43) \\  
\hline
DW+I & 48I &0.114&139&0.0048(02) &0.0198(07)& 0.0066(03)& 0.0252(05)& 0.0099(03)& 0.0198(07)\\ 
DW+I & 24I &0.111&340&0.0053(04) & 0.0164(06)& 0.0034(06) &0.0225(06)& 0.0082(08)& 0.0179(11)\\  
DW+I & 64I &0.084&139&0.0014(01) &0.0057(03)& 0.0019(04)& 0.0071(02)& 0.0028(04)& 0.0057(03)\\        
DW+I & 32I &0.083&302 &0.0015(01) &0.0055(06)& 0.0005(04)&0.0058(04) &0.0017(06) &0.0053(05)\\           
\hline
OV$^{\rm 0HYP}$+IR & JLQCD &0.112&290&0.0008(01) &0.0065(03)& 0.0024(01)& 0.0064(03)& 0.0018(01)& 0.0062(03)\\           
\hline
HI+S$^{(1)}$ & a12m310 &0.121&310&0.0371(05)& --& 0.0235(04)& 0.0371(05)& 0.0155(05)& 0.0235(04)\\  
HI+S$^{(1)}$ & a12m130 &0.121&130&0.0472(18)& --& 0.0270(31)& 0.0472(18)& 0.0229(22)& 0.0270(31)\\  
HI+S$^{(1)}$ & a09m310 &0.088&310&0.0122(06)& --& 0.0074(02)& 0.0122(06)& 0.0044(09)& 0.0074(02)\\   
HI+S$^{(1)}$ & a09m130 &0.088&130&0.0183(06)& --& 0.0078(10)& 0.0183(06)& 0.0081(08)& 0.0078(09)\\   
HI+S$^{(1)}$ & a06m310 &0.057&310&0.0023(01)& --& 0.0015(01) &0.0023(01)& 0.0018(07)& 0.0015(01)\\  
HI+S$^{(1)}$ & a04m310 &0.043&310&0.0007(03)& --& 0.0005(04) &0.0007(03)& 0.0007(01)& 0.0005(04)\\  
\hline
CL$^{\rm stout}$+S$^{\rm tad}$ & C11 &0.108&290&0.0663(35)& 0.0450(22)& 0.0103(06)& 0.0946(52)& 0.0714(34)& 0.0459(22)\\   
CL$^{\rm stout}$+S$^{\rm tad}$ & C08 &0.080&300&0.0313(27)& 0.0193(10)& 0.0051(04)& 0.0361(30)& 0.0335(25)& 0.0174(08)\\       
CL$^{\rm stout}$+S$^{\rm tad}$ & C06 &0.055&300&0.0115(18)& 0.0049(02)& 0.0020(08)& 0.0110(21)& 0.0105(19)& 0.0058(09)\\
\hline
\end{tabular}  
\end{table}

\begin{figure*}[htbp]
    \centering
    \includegraphics[width=0.45\linewidth]{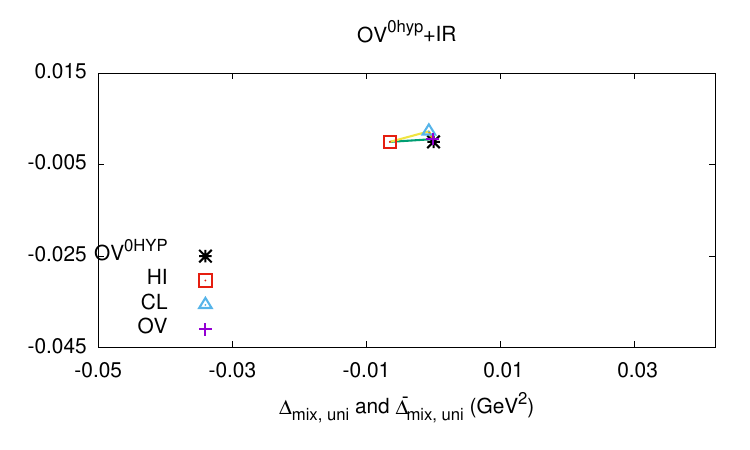}
    \includegraphics[width=0.45\linewidth]{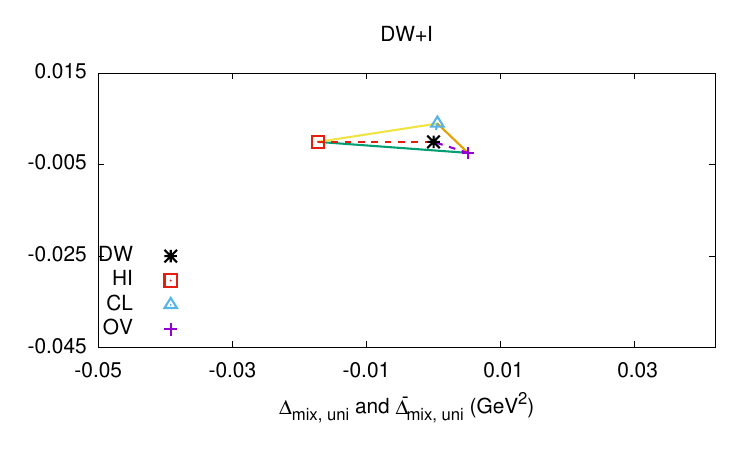}
    \includegraphics[width=0.45\linewidth]{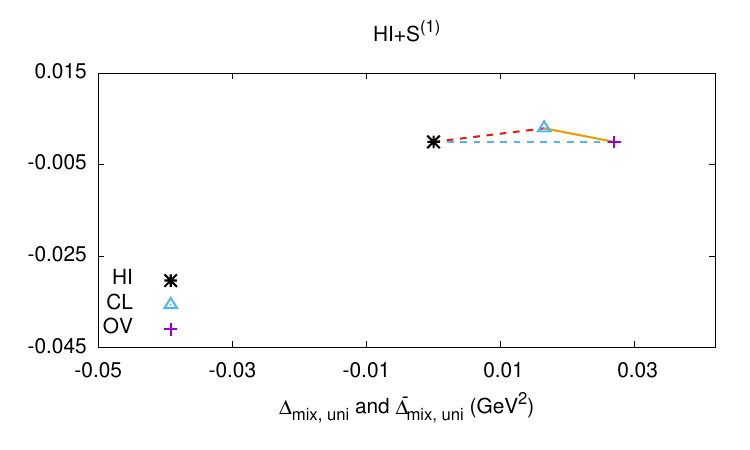}
    \includegraphics[width=0.45\linewidth]{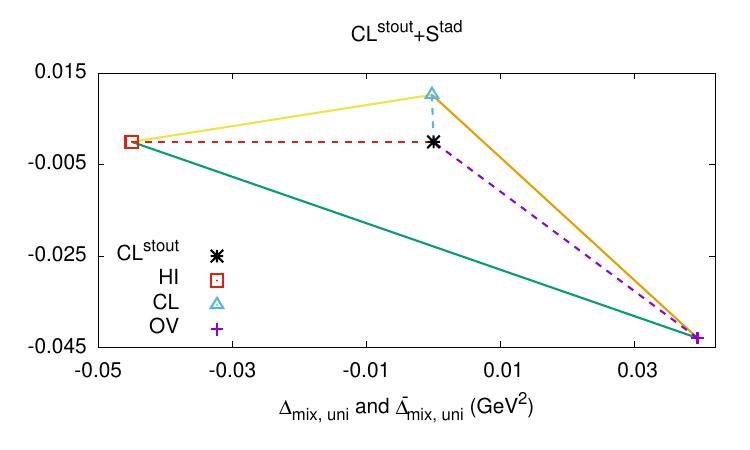}
    \caption{The mixed-action effects $\Delta_{\rm mix, uni}$ for various valence fermion actions (distances between the black star and colored data points, dashed lines) and $\bar{\Delta}_{\rm mix, uni}$ for different valence fermion actions (distances between colored data points, solid lines) on the OV$^{\rm 0HYP}$+IR (left top), DW+I (right top), HI+S$^{(1)}$ (left bottom), and CL$^{\rm stout}$+S$^{\rm tad}$ (right bottom) ensembles. All the distances are in unit of GeV$^2$, and interpolated to $a=0.11$ fm and $m_{\pi}\in [290,310]$ MeV to make a fair comparison. }
    \label{fig:triangle}
\end{figure*}

After interpolating to $a\simeq 0.11$ fm and $m_{\pi}\in [290,310]$ MeV, we show the values of $\Delta_{\rm mix, uni}$ and $\bar{\Delta}_{\rm mix, uni}$ in Fig.~\ref{fig:triangle}, on the OV$^{\rm 0HYP}$+IR, DW+I, HI+S$^{(1)}$ and CL$^{\rm stout}$+S$^{\rm tad}$ ensembles. The black star in each panel is fixed to the original point, and the distances between it and red box, blue triangle and purple crosses correspond to $\Delta_{\rm mix, uni}$ of the HI, CL and OV valence fermion actions, respectively. The distance between the colored data points corresponds to $\bar{\Delta}_{\rm mix, uni}$ of different valence fermion actions. The uncertainties of $\Delta_{\rm mix, uni}$ and $\bar{\Delta}_{\rm mix, uni}$ are smaller than 10\% in most of the cases and then ignored in the figure. 

Note that the orientation of the triangles in the sub panels of Fig.~\ref{fig:triangle} can be arbitrary, since only the distances between different actions matter. We just require the lines of $\Delta^{\rm HI/X}_{\rm mix}$ (that of $\Delta^{\rm OV/HI}_{\rm mix}$ in the HI+S$^{(1)}$ case since $\Delta^{\rm HI/HI}_{\rm mix}$ vanishes) to be along the $x$-axis, to make the triangles in all the four figures relatively similar to each other. 

One can see that the relations Eqs.~(\ref{eq:ineq_1},\ref{eq:ineq_2}) hold in all the cases, while the ``distances" between the fermion actions can be very different. The mixed-action effect on the OV$^{\rm 0HYP}$+IR ensemble (left top panel) seems to be smaller than that of the DW+I case  (right top panel), while it is comparable with the DW+ID case given the factor of 2 suppression shown in the left panel of Fig.~\ref{fig:Sea_dw_mass}. This is predictable as the DSDR term provides somehow a similar impact on the near-zero modes of the Dirac operator as does the ratio term~\cite{Fukaya:2006vs,Vranas:2006zk}. 

Compared to the mixed-action effects on the DW+I ensembles, those on the HI+S$^{(1)}$ ensemble  (left bottom panel) are somehow similar. But when we consider the valence CL or OV action, $\Delta_{\rm mix, uni}$ on the HI+S$^{(1)}$ ensemble are still larger than those on the DW+I ensemble since the HI action is somehow ``far away" from either CL, OV or DW actions. At the same time, the impact of the gauge action used by DW+I or HI+S$^{(1)}$ appears to be minor, but this requires further study using different fermion actions and the same gauge action. 

Contrary to the above three cases, the mixed-action effects on the CL$^{\rm stout}$+S$^{\rm tad}$ ensemble (right bottom panel) are much larger, even for the case of the HYP smeared clover valence fermion action on stout smeared clover sea action. The origin of the huge mixed-action effect is presumably related to the additive chiral symmetry breaking of the clover fermion action, and more investigation is essential to determine a good method to suppress it.

\end{widetext}

%
%
%

%
%

\end{document}